\documentclass[a4paper,12pt]{article}

\usepackage{amsmath}
\usepackage{amssymb}
\usepackage{mathptmx}
\usepackage{amsfonts}
\usepackage[mathscr]{eucal}

\usepackage{authblk}

\newcommand{\mbb}[1]{\ensuremath{\mathbb{#1}}}
\newcommand{\msc}[1]{\ensuremath{\mathscr{#1}}}
\newcommand{\mrm}[1]{\ensuremath{\mathrm{#1}}}

\newcommand{\Hk}{H(\kappa)}
\newcommand{\HI}{H_{\mathrm{I}}}

\newcommand{\disp}{\displaystyle}
\newcommand{\lang}{\langle}
\newcommand{\rang}{\rangle}

\newcommand{\one}{{ \mathrm{1} \hspace{-0.26em}  \mathrm{l}}}

\newtheorem{theorem}{Theorem}[section]
\newtheorem{proposition}[theorem]{Proposition}
\newtheorem{lemma}[theorem]{Lemma}
\newtheorem{corollary}[theorem]{Corollary}

\title{Non-Perturbative Analysis of Resonances by a Contraction Mapping on Complex Domains}
\author[1]{Toshimitsu Takaesu}
\author[2]{Tamotsu Teruya}
\affil[1,2]{$\,$Cooperative Faculty of Education, Gunma University, Gunma, Japan}
\date{\today}
\begin{document}

\maketitle

\begin{quote}
\textbf{[Abstract]} 
In this paper, we consider a perturbation problem of embedded eigenvalues for an abstract Friedrichs--Lee Hamiltonian and analyze its spectral properties.
By deforming the contour and constructing a contraction mapping on complex domains, we prove the existence of resonance poles.
\end{quote}


\section{Introduction and Main Results}
In this paper, we investigate an abstract   Friedrichs--Lee Hamiltonian. The Friedrichs--Lee model, originally introduced by Friedrichs \cite{Fr48} and Lee \cite{Lee54},  is a solvable model characterized by the distinctive behavior of eigenvalues embedded in the continuous spectrum.   
Our aim is to analyze the spectrum of the abstract Hamiltonian from a rigorous mathematical perspective.
 The Hilbert space for the model is defined by
$$
\msc{H} = \mbb{C} \oplus \msc{K} ,
$$
where $ \msc{K} $ is a  Hilbert space over $\mbb{C}$. Let $S$ be a linear operator on $\msc{K}$. We assume the following condition.
\begin{quote}
(A.1) $S$ is self-adjoint and non-negative. 
\end{quote}
The free Hamiltonian is given by
\begin{equation}
H_0 = 
\begin{pmatrix} \nu  & 0 \\ 0 & S \end{pmatrix} ,
\end{equation}
where $ \nu$ is a constant satisfying $ \nu >0 $.  
Let $\Omega_0 =\begin{pmatrix} 1 \\ 0  \end{pmatrix} \in \msc{D} (H_0)$.
Then $H_0 \Omega_0 = \nu \Omega_{0} $. 
Let $f \in \msc{K}$. The annihilation operator $A (f) : \msc{K} \to \mbb{C}$ is defined by $ A(f) \psi  = \lang f , \psi  \rang $. Here, the inner product is anti-linear with respect to the first argument and linear with respect to the second argument. The creation operator $A^{\dagger} (f) : \mbb{C} \to \msc{K}  $ is defined by $A^{\dagger} (f) \alpha = \alpha f$. Then, it holds that  $A^{\dagger} (f) = A(f)^{\ast}$, where $X^{\ast}$ denotes the adjoint of the operator $X$. Here, we choose a distinguished vector $g \in \msc{K} \setminus \{ 0\}$. The interaction is  defined by
\begin{equation} \HI = 
\begin{pmatrix} 0 &  A( g )  \\ A^{\dagger} (g  ) & 0 \end{pmatrix} .
\end{equation}
The total Hamiltonian is defined as
\begin{equation}
\Hk = H_{0} + \kappa \HI ,
\end{equation}
where $\kappa \in \mbb{R} $ is the coupling constant. It is seen that $\HI$ is bounded. Then, by the Kato--Rellich theorem, it follows that $\Hk $ is self-adjoint on $\msc{D} (\Hk) = \msc{D}(H_0 )$ and bounded from below.
 We assume the following conditions.
\begin{quote}
(A.2) $ \disp g \in \msc{D} (S) \cap \msc{D} ( S^{-1/2})$. \\ [2mm]
(A.3) $S$ has purely absolutely continuous spectrum with
$\sigma(S) = \sigma_{\mathrm{ac}}(S) = [0, \infty) $.
\end{quote}

From the spectral properties of the direct sum of linear operators, it holds that $\sigma (H_0) = \lbrack 0 , \infty ) $ and $\sigma_{\mrm{p}} (H_0) = \{ \nu \}$.
Since $\nu $ is an eigenvalue embedded in the continuous spectrum, its stability under the interaction is not guaranteed. In fact, as we will see later, this eigenvalue disappears and a resonance pole appears for  sufficiently small $| \kappa | \ne 0 $.  Here, we briefly consider the essential spectrum of $\Hk $.
Let $\Psi = \begin{pmatrix} \alpha \\ \psi  \end{pmatrix} \in \msc{H}$. Then, we have $\HI \Psi = \begin{pmatrix} \lang g , \psi \rang \\ \alpha g   \end{pmatrix} $. This implies that  
$\mrm{Ran} \, \HI = \mrm{Span}\, \left\{ \begin{pmatrix} 1 \\ 0 \end{pmatrix}, \begin{pmatrix} 0 \\ g \end{pmatrix}  \right\}$.
Therefore, $\HI $ is a finite-rank operator and thus compact. Hence, by Weyl's theorem,  it holds that 
\begin{equation}
\sigma_{\mrm{ess}} (\Hk ) = \sigma_{\mrm{ess}} (H_{0}) = \lbrack 0 ,\infty ).
\end{equation}

Let  $E_S$ be the spectral projection of $S$. The spectral measure associated with $S$ and $f , f' \in \msc{K}$ is denoted by $ \mu^{S}_{f, f' } (J) = \lang f , E_S (J) f' \rang $, where $J \subset [ 0, \infty )$ is a Borel set. From (A.3) and the Radon--Nikodym theorem,  there exists a density function $\rho^{S}_{f, f'} $ satisfying $d\mu^{S}_{f, f'} (\lambda ) = 
\rho_{f, f'}^{S} (\lambda ) d \lambda  $. 
For simplicity, we set $ \mu^{S}_{f}  =\mu^{S}_{f, f} $ and $\rho^{S}_{f} = \rho^{S}_{f , f}   $. 
We impose the following condition.
\begin{quote}
(A.4) $\rho^{S}_{g} $ is continuous on $[0, \infty)$, and  $\rho^{S}_{g} (\lambda)> 0$ for all $\lambda \in (0, \infty) $. 
\end{quote}

\begin{proposition} \label{non-existEV}
Assume (A.1)--(A.3). \\[1mm]
(i) If (A.4) holds,    $\Hk$ has no eigenvalues  in $( 0, \infty )$ for   all $\kappa \ne 0 $. \\
(ii) $\Hk $ has no  zero eigenvalue for all $\kappa$ satisfying $0<  |\kappa | \ne \frac{\sqrt{\nu}}{\| S^{-1/2} g \|}$. \\
(iii) $\Hk $  has no  eigenvalues in $ ( - \infty , 0 )$  for all $\kappa $  satisfying  $0< |\kappa| <  \frac{\sqrt{\nu}}{\| S^{-1/2} g \|}$.  \\
In particular, under (A.1)--(A.4), $\Hk $ has no eigenvalues  for   all $\kappa $  satisfying  $0< |\kappa| <  \frac{\sqrt{\nu}}{\| S^{-1/2} g \|}$.
\end{proposition}
Note that $\| S^{-1/2} g \|>0$ since $g \ne 0$. 
On the other hand, the following holds. 

\begin{proposition} \label{ExistEV}
Assume (A.1)--(A.3). For all $\kappa $  satisfying  $ |\kappa| > \frac{\sqrt{\nu}}{\| S^{-1/2} g \|}$, $\Hk $  has  a unique negative  eigenvalue. 
\end{proposition}
Thus, under (A.1)--(A.4), we see that $\Hk$ has no eigenvalues when $|\kappa| \neq 0$ 
is smaller than a certain value, whereas it has a unique negative eigenvalue when $|\kappa|$ is larger than that value.

Here we define resonance poles. 
Let $H$ be a self-adjoint operator on a Hilbert space.
If there exist vectors $\Phi$ and $\Psi $ such that
 $ f_{\Phi, \Psi}(z) = \lang \Phi, (H-z)^{-1} \Psi \rang$ has a meromorphic continuation $\tilde{f}_{\Phi, \Psi}(z)$ from the upper half-plane $\mbb{C}_{+}= \{ z \in \mbb{C} \mid \mrm{Im} \, z > 0 \} $ across an interval of the real axis into a domain $D$ in  the lower half-plane $\mbb{C}_{-}= \{ z \in \mbb{C} \mid \mrm{Im} \, z < 0 \} $, and $\tilde{f}_{\Phi, \Psi}(z)$ has a pole in $D$, then $H$ is said to have a resonance pole. 
For $\alpha \in \mbb{C}$ and $r>0$,  we set $D_{r}(\alpha ) = \{ z \in \mbb{C} \mid |z-\alpha | < r \}$.  For $0< s  < \nu $,  we define $I_{s} = (\nu -s  , \nu + s )$.
Assume the following additional condition regarding the density 
function $\rho^{S}_{g}$.
\begin{quote}
(A.5) There exists $0 < \delta < \nu $  such that $\rho^{S}_{g}$ is analytic on $I_{\delta} $ and has an analytic continuation $\tilde{\rho}^{S}_{g}$ to $D_{\delta}(\nu )$.
\end{quote}
Resonances in the Friedrichs--Lee model have been mathematically 
investigated from various viewpoints. For details, we refer to 
\cite{King91} for exponential decay near resonances, 
\cite[Chapter 5]{Arai05} for  the analysis of resonances  in  the concrete  model,  \cite{FLL21} for that in  the singular model, and the references therein.  Additionally, a comprehensive review of various  Friedrichs models is given in \cite{GP11}.

For $z \in \mbb{C} \setminus \mbb{R}$, let $f_{\Omega_0}(z) = \lang \Omega_0, (\Hk - z)^{-1} \Omega_0 \rang$. It is seen that 
$$ f_{\Omega_0}(z) = \frac{1}{F_{\kappa} (z) }
$$
 where $F_{\kappa} (z) = \nu - z -\kappa^2 \Sigma_g (z)$ with $\Sigma_g (z) = \lang g , (S-z)^{-1} g \rang $. 
We investigate the  meromorphic continuation of $f_{\Omega_0}(z) $.   In quantum mechanics, it is known that the method of complex 
dilation \cite{AgCo71, BaCo71} is effective for constructing 
meromorphic continuations (see, e.g., \cite[Chapter 16]{HiSi96}, 
\cite[Chapter XII]{ReSi78}). This approach has achieved 
significant success in quantum field theory \cite{BFS99} as well. 
However, while the method of complex dilation can be applied to 
concrete models, it is not applicable to the abstract 
Friedrichs--Lee model, since no dilation structure is assumed on the abstract Hilbert space $\msc{K}$.  
In this paper, we  apply  the method of contour deformation.
Let $\delta \in (0, \nu)$ be a constant in (A.5), and let $0 < r < \delta$. Then we choose $R>0$ such that  $r < R< \delta$. We show that $\Sigma_g(z)$ has an 
analytic continuation $\tilde{\Sigma}_g(z)$ from 
$\mbb{C}_{+}$ across $I_{R}$ into 
$D_{R}(\nu) \cap \mbb{C}_{-}$. Let $\tilde{f}_{\Omega_{0}}(z) = \frac{1}{\tilde{F}_{\kappa}(z)}$, where 
$\tilde{F}_{\kappa}(z) = \nu - z - \kappa^2 \tilde{\Sigma}_g(z)$. 
Since $0 < r < R$, we see that $\overline{D}_{r}(\nu) \subset D_{R}(\nu)$, where 
$\overline{D}$ denotes the closure of $D \subset \mbb{C}$. 
Then,  $\tilde{F}_{\kappa}(z)$ is analytic on $\overline{D}_{r}(\nu)$. 
We analyze  $\tilde{F}_{\kappa}(z)$  on $\overline{D}_{r}(\nu)$ and prove  that $\Hk$ has a resonance pole as follows.

\begin{theorem} \label{MainTheorem}
Assume (A.1)--(A.5), and let $0 < r < \delta $.  Then, there exists some $\kappa_{\ast}>0$ such that for all $\kappa$ satisfying $0< |\kappa| < \kappa_{\ast}$,  $\tilde{f}_{\Omega_0}(z) $ has a unique pole  in  $\overline{D}_{r}(\nu) \cap \mbb{C}_{-}$. 
\end{theorem}

\noindent
The key idea of the proof is that finding a pole of 
$\tilde{f}_{\Omega_0}(z)$ in $\overline{D}_{r}(\nu)$ is equivalent to finding a fixed point 
of the map 
\[
G_{\kappa}: z \mapsto \nu - \kappa^2 \tilde{\Sigma}_{g}(z) 
\]
 in $\overline{D}_{r}(\nu)$. We prove the existence and 
uniqueness of the fixed point by the contraction mapping 
principle and  show that  it lies in $\mbb{C}_{-}$.

From Proposition \ref{non-existEV} and Theorem \ref{MainTheorem},  we see that  for all  $\kappa$ satisfying $0< | \kappa | < \min \left\{   \frac{\sqrt{\nu}}{\| S^{-1/2} g \|} , \kappa_{\ast} \right\} $, $\Hk$ has no eigenvalues but possesses a resonance pole.

 Here, we consider  $f_{\Phi, \Psi}(z) = \lang \Phi, (\Hk - z)^{-1} \Psi \rang$  for general vectors $\Phi, \Psi \in \msc{H}$ and 
$z \in \mbb{C} \setminus \mbb{R}$.
Let $\Phi = \begin{pmatrix} \alpha \\ \phi \end{pmatrix}, 
\Psi = \begin{pmatrix} \beta \\ \psi \end{pmatrix} \in \msc{H}$. 
By Corollary \ref{stateResolvent}, it holds that for  all
$z \in \mbb{C} \setminus \mbb{R}$,
\begin{equation}
f_{\Phi, \Psi}(z) = \frac{1}{ F_{\kappa} (z)}
\left( \overline{\alpha} - \kappa \Sigma_{\phi, g}(z) \right) 
\left( \beta - \kappa \Sigma_{g, \psi}(z) \right) 
+ \Sigma_{\phi, \psi}(z) ,
\end{equation}
where $\Sigma_{f, f'}(z) = \lang f, (S - z)^{-1} f' \rang$ 
for $f, f' \in \msc{K}$. For given $\phi, \psi \in \msc{K}$, 
we introduce the following assumption.

\begin{quote}
(A.6) Let $ 0 < \delta < \nu$ be a constant satisfying (A.5). The functions  
$\rho^{S}_{\phi, g}$, $\rho^{S}_{g, \psi}$, and 
$\rho^{S}_{\phi, \psi}$ are analytic on $I_{\delta}$, and they have 
analytic continuations $\tilde{\rho}^{S}_{\phi, g}$, 
$\tilde{\rho}^{S}_{g, \psi}$, and $\tilde{\rho}^{S}_{\phi, \psi}$ 
to $D_{\delta}(\nu)$, respectively. 
\end{quote}

\noindent
Let $0 < r < \delta$. Similarly to $f_{\Omega_0}(z)$, under (A.1)--(A.6), 
 for all $R>0$ satisfying $r <R < \delta$, $f_{\Phi, \Psi}(z)$ has a meromorphic 
continuation $\tilde{f}_{\Phi, \Psi}(z)$ from $\mbb{C}_{+}$ across 
$I_{R}$ into $D_{R}(\nu) \cap \mbb{C}_{-}$, given by
\begin{equation}
\tilde{f}_{\Phi, \Psi}(z) = \frac{1}{\tilde{F}_{\kappa} (z) } 
\left( \overline{\alpha} - \kappa \tilde{\Sigma}_{\phi, g}(z) \right) 
\left( \beta - \kappa \tilde{\Sigma}_{g, \psi}(z) \right) 
+ \tilde{\Sigma}_{\phi, \psi}(z) ,
\end{equation}
where $\tilde{\Sigma}_{f, f'}(z)$ denotes the analytic continuation 
of $\Sigma_{f, f'}(z)$ to $D_{R}(\nu)$ constructed by contour deformation.  Then,  we  analyze $\tilde{f}_{\Phi, \Psi}(z)$ on  $ \overline{D}_{r}(\nu) \cap \mbb{C}_{-}$ and the following holds. 

\begin{corollary} \label{Main-coro}
Assume (A.1)--(A.5), and let $0 < r < \delta$. Let $\Phi = \begin{pmatrix} \alpha \\ \phi \end{pmatrix}, 
\Psi = \begin{pmatrix} \beta \\ \psi \end{pmatrix} \in \msc{H}$ be vectors such that $\alpha \neq 0$, $\beta \neq 0$, and $\phi, \psi$ satisfy (A.6). 
Then, there exists $0< \kappa_{0} \leq  \kappa_{\ast}$ such that for all $\kappa$ satisfying $0 <|\kappa|  <  \kappa_{0}$, $\tilde{f}_{\Phi, \Psi}(z)$ 
has a unique pole in $ \overline{D}_{r}(\nu) \cap \mbb{C}_{-}$, which is the pole given in Theorem \ref{MainTheorem}.
\end{corollary}
\textit{Proof.}  Let $r < R < \delta$.
 Since $\tilde{\Sigma}_{\phi, g}(z)$ and 
$\tilde{\Sigma}_{g, \psi}(z)$ are analytic on ${D}_{R}(\nu)$,
we can set $M_1 = \disp \max_{z\in \overline{D}_{r}(\nu) }|\tilde{\Sigma}_{\phi, g}(z) | < \infty $ and   $M_2 = \disp \max_{z \in \overline{D}_{r}(\nu) }|\tilde{\Sigma}_{g, \psi}(z) | < \infty $.  Since $\alpha \neq 0$ and $\beta \neq 0$, we can take $\kappa_1 > 0$ such that 
$ | \alpha |  - \kappa_1 M_1 >0 $ and $|\beta |  - \kappa_1 M_2  >0 $.
Then, for all $\kappa$  satisfying $ 0 < | \kappa |< \kappa_1 $, it holds  that $| \overline{\alpha} - \kappa \tilde{\Sigma}_{\phi, g}(z) | \ne 0  $ and 
$| \beta - \kappa \tilde{\Sigma}_{g, \psi }(z) |  \ne 0 $ for all $z \in \overline{D}_{r}(\nu) $. Furthermore, 
$\tilde{\Sigma}_{\phi, \psi}(z)$ is analytic on $D_{R} (\nu)$. 
Let $\kappa_0 = \min \left\{ \kappa_1 , \kappa_{\ast} \right\}$. Then,  by Theorem \ref{MainTheorem}, $\tilde{f}_{\Phi, \Psi}(z)$ has a unique pole in $\overline{D}_{r}(\nu) \cap \mbb{C}_{-}$ for all $\kappa$ satisfying $0 < | \kappa | < \kappa_0 $. $\square$ \\

This paper is organized as follows. In Section 2, 
we prove the non-existence and existence of eigenvalues. 
In Section 3, we derive the explicit form of the resolvent, 
construct the analytic continuation by contour deformation, 
and prove the existence of a resonance pole. 
In Section 4, we give concrete examples satisfying the assumptions.


\section{Eigenvalues}
\subsection{Proof of Proposition \ref{non-existEV}} 
First we show the following lemma. 

\begin{lemma} \label{divergence}
Assume (A.1)--(A.4). Then, for all $\gamma >0$, $ \disp \| (S- \gamma)^{-1} g \|=  \infty $ holds. 
\end{lemma}
\textit{Proof.} Let $\gamma > 0 $. 
By the spectral theorem and (A.3), we have
$$
 \| (S-\gamma )^{-1} g \|^2  
 = \int_{0}^{\infty} \frac{1}{( \lambda - \gamma )^2} d \mu^{S}_{g} (\lambda )
 =\int_{0}^{\infty} \frac{\rho^{S}_{g} (\lambda)}{( \lambda - \gamma )^2} d\lambda .
 $$
By (A.4),   $\rho^{S}_{g} $ is continuous on $[0, \infty)$ and  
$\rho^{S}_{g} (\lambda)>0$ for all $\lambda >0$. Then, we obtain $\disp \int_{0}^{\infty} \frac{\rho_{g}^{S} (\lambda )}{( \lambda - \gamma )^2} d\lambda = \infty $. 
Therefore, $\| (S - \gamma)^{-1} g \| = \infty$. $\square$ \\

$\;$ \\
\textit{Proof of Proposition \ref{non-existEV}.} \\
Suppose that $\Hk \Psi = E \Psi $ where  $E \in \mathbb{R}$ and  $\Psi = \begin{pmatrix} \alpha \\ \psi \end{pmatrix}\in \msc{D}(\Hk  )\setminus \{ 0 \} $. Then,  we  have 
\begin{align}
& (\nu - E) \alpha  = - \kappa \lang g , \psi \rang  , \label{E1a} \\
& (S-E) \psi = - \kappa \alpha  g . \label{E1b}
\end{align}
We show that $\alpha  \ne 0$ by contradiction.
Suppose that $\alpha  =0 $.  By  (\ref{E1b}),  we have $S \psi = E \psi $. By (A.3),  $\sigma_{\mrm{p}} (S) = \emptyset $ holds. Hence, $\psi =0 $, which implies $\Psi = 0$. This contradicts $\Psi \ne 0 $. Therefore, $\alpha  \ne 0 $.
We prove  (i)--(iii) as follows. \\
(i) Assume (A.4) and $\kappa \ne 0$.  Suppose $E > 0$. From (\ref{E1b}), we see that $g \in \mathrm{Ran} (S-E)$. Since $\sigma_{\mathrm{p}} (S) = \emptyset$, $ S-E $ is injective. Since $\kappa  \ne 0 $ and $\alpha \ne 0 $, we have
$$
\|  (S-E )^{-1} g \| \leq \frac{1}{|\kappa| | \alpha  |} \|  \psi \| < \infty.
$$
This contradicts Lemma \ref{divergence}. 
Therefore, $\Hk$ has no eigenvalues in $(0, \infty)$.\\
(ii) Assume $ |\kappa| \ne  \frac{\sqrt{\nu}}{\| S^{-1/2} g \|}$. Suppose  $E=0$.  From (\ref{E1b}), we have $S \psi = - \kappa \alpha  g $. Then we obtain
\begin{equation}
S^{1/2} \psi = - \kappa \alpha  S^{-1/2}  g  . \label{E1c}
\end{equation}
From (\ref{E1a}) and (\ref{E1c}), we have
$$
\nu \alpha  = - \kappa \lang S^{-1/2} g ,  S^{1/2}  \psi   \rang 
=  \kappa^2 \alpha   \|   S^{-1/2} g \|^2 . 
$$
Since $\alpha  \ne 0$, we have $\nu  = \kappa^2 \|   S^{-1/2} g \|^2$.
This contradicts  $|\kappa| \ne  \frac{\sqrt{\nu}}{\| S^{-1/2} g \|}$. Hence, $\Hk$ has no zero eigenvalue. \\
(iii) Assume $0< |\kappa| <  \frac{\sqrt{\nu}}{\| S^{-1/2} g \|}$. 
Suppose  $E<0$.  Since $ \sigma (S) = [ 0 , \infty) $, $(S-E)^{-1}$ is bounded. From (\ref{E1b}), we have $ \psi =- \kappa \alpha  (S-E )^{-1}   g$.  Substituting this into (\ref{E1a}) and using $\alpha \neq 0$, we obtain
$$
E =\nu - \kappa^2 \lang g  , (S-E)^{-1} g \rang .
$$
By the spectral theorem, we obtain 
\begin{equation}
E =\nu - \kappa^2 \int_0^{\infty} \frac{1}{ \lambda -E } d \mu^{S}_{g} (\lambda) . \label{E1d}
\end{equation}
Here, the right-hand side of (\ref{E1d}) is evaluated as
\begin{align*}
\nu - \kappa^2 \int_0^{\infty} \frac{1}{\lambda - E} d \mu^{S}_{g }(\lambda) 
& > \nu - \kappa^2 \int_0^{\infty} \frac{1}{\lambda} d \mu^{S}_{g}(\lambda) \\
& = \nu - \kappa^2 \| S^{-1/2} g \|^2 > 0 ,
\end{align*}
where the last inequality follows from  $|\kappa| <  \frac{\sqrt{\nu}}{\| S^{-1/2} g \|}$. 
On the other hand, the left-hand side of (\ref{E1d}) satisfies $E < 0$, 
which yields a contradiction. Therefore, $\Hk$ has no eigenvalues 
in $(-\infty, 0)$. This completes the proof of Proposition \ref{non-existEV}. 
 $\square $\\

\subsection{Proof of Proposition \ref{ExistEV}}
Suppose that $\Hk \Psi = E \Psi$, where $E < 0$ and 
$\Psi = \begin{pmatrix} \alpha \\ \psi \end{pmatrix} \in \msc{D}(\Hk) \setminus \{ 0 \}$. 
In a similar way to the proof of Proposition \ref{non-existEV}, 
it follows that $\alpha \ne 0$ and $
E = \nu - \kappa^2 \int_0^{\infty} \frac{1}{\lambda - E} \, d\mu^{S}_{g}(\lambda)$.
We define a function $f$ on $(-\infty , 0)$ by
$$
f(E) = \nu - \kappa^2 \int_0^{\infty} \frac{1}{\lambda - E} \, d\mu^{S}_{g}(\lambda) - E .
$$
We see that $f(E)$ is continuous and strictly decreasing on 
$(-\infty, 0)$, $\lim\limits_{E \to -\infty} f(E) = \infty$, and 
$\disp \lim\limits_{E \to 0-} f(E) = \nu - \kappa^2 \| S^{-1/2} g \|^2 < 0$ 
since $|\kappa| > \frac{\sqrt{\nu}}{\| S^{-1/2} g \|}$. Hence, there exists 
a unique $E_0 < 0$ such that $f(E_0) = 0$. 
Conversely, for this $E_0$, we set $
\Psi_0 = \begin{pmatrix} 1 \\ -\kappa (S - E_0)^{-1} g \end{pmatrix} $.
Note that  $(S - E_0)^{-1} g \in \msc{D}(S)$, since $E_0 < 0$ and $\sigma(S) = [0, \infty)$. 
By a direct calculation using $f(E_0) = 0$,  
we have  
$\Hk \Psi_0 = E_0 \Psi_0$. Thus, $E_0$ is the unique 
negative eigenvalue of $\Hk$. This completes the proof of Proposition \ref{ExistEV}. $\square$

\section{Resonances}

\subsection{Resolvent}
The resolvent of $\Hk$ is explicitly obtained as follows.
\begin{lemma} \label{exactResolvent}
Assume (A.1) and (A.2). Then, for all $z \in \mbb{C} \setminus  \mbb{R}$, 
$$
(\Hk -z)^{-1} =
\begin{pmatrix}
 \frac{1}{F_{\kappa} (z) } & -  \frac{ \kappa  }{F_{\kappa} (z) }  A \left( (S-\overline{z} )^{-1} g \right) \\
 -    \frac{\kappa }{F_{\kappa} (z) }   A^{\dagger} \left( (S-z)^{-1} g \right)
 & (S-z)^{-1}  +
  \frac{\kappa^2 }{F_{\kappa} (z) } A^{\dagger} \left( (S-z)^{-1} g \right)  A \left( (S-\overline{z} )^{-1} g \right) 
   \end{pmatrix} ,
 $$
 where $F_{\kappa} (z) = \nu - z - \kappa^2 \lang g , (S-z)^{-1} g \rang $.
 \end{lemma}
\textit{Proof.} Let $\Psi =  \begin{pmatrix}  \alpha \\ \psi \end{pmatrix} \in \msc{H}$ and $\Phi = \begin{pmatrix}  \beta \\ \phi \end{pmatrix} \in \msc{D}(\Hk  )$. Suppose that $ (\Hk -z)^{-1} \Psi = \Phi $. Then, $\Psi = (\Hk -z) \Phi $, and the following holds:
\begin{align}
& \alpha = ( \nu -z ) \beta + \kappa A( g  ) \phi ,   \label{Res1a} \\ 
& \psi  = \kappa A^{\dagger}(g) \beta  + (S-z) \phi \label{Res1b}  .
\end{align}
By (\ref{Res1b}) and $ (S-z)^{-1} A^{\dagger } (g ) =A^{\dagger }  \left(  (S-z)^{-1}  g \right)$, we have  
\begin{equation}
\phi  = (S-z)^{-1} \psi  - \kappa   A^{\dagger} \left( (S-z)^{-1}g \right)    \beta \label{Res1c}  .
\end{equation}
Substituting  (\ref{Res1c}) into (\ref{Res1a}), and using the relations
$ A (g ) (S-z)^{-1}  =A  \left(  (S-\overline{z})^{-1}  g \right)$ and 
 $A(g )  A^{\dagger} \left( (S-z)^{-1} g \right) = \lang g , (S-z)^{-1} g \rang$, we obtain 
\begin{equation}
\beta =    \frac{1}{F_{\kappa} (z) } \left( \alpha  -\kappa   A \left( (S-\overline{z} )^{-1} g \right)  \psi  \right) , \label{Res1d}
\end{equation}
where $F_{\kappa} (z) = \nu -z - \kappa^2 \lang g , (S-z)^{-1} g \rang $.
Substituting (\ref{Res1d}) into (\ref{Res1c}), we obtain
\begin{equation}
\phi =  - \frac{\kappa}{F_{\kappa} (z)} A^{\dagger } ( (S-z)^{-1} g ) \alpha
+\left(  (S-z)^{-1}   + \frac{\kappa^2}{F_{\kappa} (z)}   A^{\dagger } \left( (S-z)^{-1} g \right) A\left( (S-\overline{z})^{-1} g \right) \right)  \psi .  \label{Res1e}
\end{equation}
From (\ref{Res1d}) and (\ref{Res1e}), the assertion holds. $\square$

\begin{corollary} \label{stateResolvent}
Assume (A.1) and (A.2). Let $\Phi =  \begin{pmatrix}  \alpha  \\
 \phi  \end{pmatrix} , \, \Psi  =  \begin{pmatrix}  \beta  \\
 \psi  \end{pmatrix} \in \msc{H}$. Then, for all $z \in \mbb{C} \setminus  \mbb{R}$, 
$$
\lang \Phi , (\Hk -z)^{-1} \Psi \rang
 = \frac{1}{ F_{\kappa} (z)}
\left( \overline{\alpha} - \kappa 
\lang \phi , (S-z)^{-1} g \rang  \right) 
\left( \beta - \kappa  \lang g , (S-z)^{-1} \psi  \rang \right) 
+  \lang \phi , (S-z)^{-1} \psi  \rang  .
$$
\end{corollary}

\subsection{Analytic Continuation by Contour Deformation}
Throughout this section, we assume (A.1)--(A.5).
From Corollary \ref{stateResolvent}, it holds that 
$$
\lang \Omega_0  , (\Hk -z)^{-1} \Omega_0 \rang = \frac{1 }{F_{\kappa}(z)}  ,
$$ where $
F_{\kappa} (z) = \nu - z - \kappa^2 \lang g , (S-z)^{-1} g \rang  $. 
Let
$$
\Sigma_g (z) = \lang g , (S-z)^{-1} g \rang .
$$
To construct the meromorphic continuation of 
$f_{\Omega_0 }(z) = \lang \Omega_0, (\Hk - z)^{-1} \Omega_0 \rang$ 
from $\mbb{C}_{+}$ into $\mbb{C}_{-}$ across a neighborhood of $\nu$ 
on the real axis, it suffices to construct that of $\Sigma_{g}(z)$.
Let $0< \delta  < \nu $ be a constant satisfying (A.5); $\rho^{S}_{g}$ is analytic on $I_{\delta} = (\nu - \delta, \nu + \delta)$ and has an analytic continuation $\tilde{\rho}^{S}_{g}$ to $D_{\delta}(\nu)$. Let 
$0 <  R < \delta$. By the spectral  theorem and by (A.3),  $ \Sigma_{g} (z)  $ can be expressed as
\begin{equation}
\Sigma_{g} (z) 
= \int_{[0, \infty) \setminus I_{R}} \frac{1}{\lambda -z } 
d \mu^{S}_{g}(\lambda ) + \int_{I_{R}} \frac{\rho^{S}_{g} (\lambda )}{\lambda -z} d \lambda
, \quad z \in\mbb{C} \setminus \mbb{R} . 
\label{sigma} 
\end{equation}
Here we deform the integration contour of the second term on the right-hand side into the lower half-plane as follows.
Let $C_{R} = \{ \zeta \in \mbb{C} \mid \zeta = \nu + R e^{i \theta} , \pi \leq  \theta \leq 2 \pi \}$. For all $z \in D_{R}(\nu )$, we define 
\begin{equation}
\tilde{\Sigma}_g (z)
= \int_{[0, \infty) \setminus I_{R}} \frac{1}{\lambda -z } d \mu^{S}_{g}(\lambda ) + \int_{C_{R}} \frac{ \tilde{\rho}^{S}_{g} (\zeta )}{\zeta -z} d \zeta  .
\label{tildesigma}
\end{equation}
It is seen that $\tilde{\Sigma}_{g}(z)$ is analytic on $D_{R}(\nu)$.
For all $z \in D_{R}(\nu )  \cap \mbb{C}_{+} $, by Cauchy's integral theorem, we have
\begin{equation}
 \int_{C_{R}} \frac{\tilde{\rho}^{S}_{g} (\zeta )}{\zeta -z} d \zeta - \int_{I_{R}} \frac{\rho^{S}_{g} (\lambda )}{\lambda -z} d \lambda = 0 . \label{11/1a}
\end{equation}
Substituting (\ref{11/1a}) into (\ref{tildesigma}), we obtain 
$$\tilde{\Sigma}_{g} (z) = \Sigma_{g} (z).$$
 On the other hand, for all $z \in D_{R}(\nu )  \cap \mbb{C}_{-}$, by Cauchy's integral formula, we have
 \begin{equation}
\int_{C_{R}} \frac{\tilde{\rho}^{S}_{g} (\zeta )}{\zeta -z} d \zeta - \int_{I_{R}} \frac{\rho^{S}_{g} (\lambda )}{\lambda -z} d \lambda  = 
2 \pi i \tilde{\rho}^{S}_{g} (z)  . \label{11/1b}
\end{equation}
Substituting (\ref{11/1b}) into (\ref{tildesigma}), we obtain
$$
\tilde{\Sigma}_g (z) = \Sigma_g (z) + 2 \pi i \tilde{\rho}^{S}_{g} (z) . 
$$

\subsection{Proof of Theorem \ref{MainTheorem}}
Let $0 < r < \delta$, and choose $R$ such that $r < R < \delta$. 
Recall that $f_{\Omega_0 }(z) = \frac{1}{F_{\kappa}(z)}$ where
$F_{\kappa}(z) = \nu - z - \kappa^2 \Sigma_g (z)$ for $z \in \mbb{C} \setminus \mbb{R}$.  Let $\tilde{\Sigma}_g(z)$ be the analytic continuation of $\Sigma_g(z)$ 
to $D_{R}(\nu)$ constructed in Section 3.2. We define 
$\tilde{F}_{\kappa}(z) = \nu - z - \kappa^2 \tilde{\Sigma}_g(z)$ on  $ D_{R}(\nu)$, and set
$$
\tilde{f}_{\Omega_0 }(z) = \frac{1}{\tilde{F}_{\kappa}(z)} .
$$
To prove that $\Hk$ has a resonance pole in $\overline{D}_{r}(\nu) \cap \mbb{C}_{-}$, we show that $\tilde{F}_{\kappa}(z)$ 
has a zero in $\overline{D}_{r}(\nu) \cap \mbb{C}_{-}$.
 Let
$$
G_{\kappa}(z) = \nu  - \kappa^2 \tilde{\Sigma}_g (z) .
$$
We prove that $G_{\kappa}$ has a fixed point in  
$\overline{D}_{r} (\nu) $, by the Banach fixed-point  theorem.
 First, we show that $G_{\kappa}$ maps $\overline{D}_{r} (\nu)$ into $\overline{D}_{r} (\nu)$.  Let $ \disp L_{r} = \max_{z \, \in  \, \overline{D}_{r}(\nu) } \left| \tilde{\Sigma}_g (z) \right|$. Then, we have
  \begin{equation}
 | G_{\kappa}(z) - \nu  | = \kappa^2 \left|
 \tilde{\Sigma}_g (z)   \right|
 \leq  \kappa^2 L_{r} ,
 \end{equation}
 for all $z \in \overline{D}_r(\nu) $. If $| \kappa | \leq  \sqrt{\frac{r}{L_{r}}}$, we  have  $ | G_{\kappa}(z) - \nu  | \leq r$. Next, we show that $G_{\kappa}$ is a contraction mapping. Let $ \disp M_{r} = \max_{z \, \in \, \overline{D}_{r} (\nu) }\left| \tilde{\Sigma}_{g}' (z) \right|$. 
 Since  $\overline{D}_{r}(\nu) $ is convex and $\tilde{\Sigma}_{g}  $ is analytic on $\overline{D}_{r}(\nu) $, the mean value inequality yields, for all $z_1 , z_2 \in \overline{D}_{r}(\nu) $,
 \begin{equation}
  | G_{\kappa}(z_1) -   G_{\kappa}(z_2) | =  
  \kappa^2  \left|  \tilde{\Sigma}_g (z_1 )  - \tilde{\Sigma}_g (z_2 )   \right|      \leq \kappa^2 M_{r} |z_1 - z_2| .
 \end{equation}
If $| \kappa | <   \frac{1}{\sqrt{M_r }}$, we have $\kappa^2 M_r < 1$. 
Let $\kappa_{\ast} = \min \left\{ \sqrt{\frac{r}{L_{r}}} , \frac{1}{\sqrt{M_r }} \right\}$.  
Note that $L_r > 0$ and $M_r > 0$ since $\tilde{\Sigma}_g$ is not constant by (A.4) and $g \ne 0$.
Then, for $|\kappa| < \kappa_{\ast}$, by the Banach fixed-point theorem, there exists a unique $z_\kappa \in \overline{D}_{r}(\nu)$ such that $G_{\kappa}(z_\kappa  ) =z_\kappa $.

Finally, we  show that $\tilde{F}_{\kappa} $ has a zero in $ \overline{D}_{r}(\nu) \cap \mbb{C}_{-}$. For $z\in \overline{D}_{r}(\nu)  \cap \mbb{C}_{+}$,  $ \tilde{F}_{\kappa} (z) = F_{\kappa} (z)$. Since $F_{\kappa}(z)$ has no zeros in $\mbb{C}_{+}$,  $\tilde{F}_{\kappa} (z)$ has no zeros in $\overline{D}_{r}(\nu) \cap \mbb{C}_{+}$. 
  Let $z= x \in [\nu - r, \nu + r]$. Then, 
$$
\tilde{\Sigma}_g (x)
= \int_{[0, \infty) \setminus I_{R}} \frac{1}{\lambda -x} d \mu^{S}_{g}(\lambda ) + \int_{C_{R}} \frac{ \tilde{\rho}^{S}_{g} (\zeta )}{\zeta -x} d \zeta  .
$$
We set
$I (x )=\int_{C_{R}} \frac{ \tilde{\rho}^{S}_{g} (\zeta )}{\zeta -x} d \zeta $. By the Schwarz reflection principle, we have $\overline{I(x)} = \int_{C_{R}} \frac{ \tilde{\rho}^{S}_{g} (\overline{\zeta} )}{\overline{\zeta} -x} d \overline{\zeta} $.
Let $C_{R}^{+} = \{ \zeta \in \mbb{C} \mid \zeta = \nu + R e^{i \theta} ,  0 \leq  \theta \leq  \pi \}$.
Since $ \overline{I(x)} =-\int_{C_{R}^{+}} \frac{ \tilde{\rho}^{S}_{g} (\zeta )}{\zeta -x} d \zeta $, by Cauchy's integral formula, we have
$$
I(x) -\overline{I(x)} = \int_{C_{R} \cup C_{R}^{+}} 
\frac{ \tilde{\rho}^{S}_{g} (\zeta )}{\zeta -x} d \zeta = 2 \pi i 
\rho^{S}_{g} (x ) .
$$
Thus, we obtain $\mrm{Im} \,  I (x)=\pi \rho^{S}_{g} (x ) $. From (A.4),  
$\rho^{S}_{g} (x )\ne 0$, and  hence, $\mrm{Im} \, I (x) \ne 0$. Since $ \tilde{F}_{\kappa} (x) = \nu -x -  \kappa^2 \tilde{\Sigma}_{g} (x) $, we obtain $ \mrm{Im} \, \tilde{F}_{\kappa} (x) = - \kappa^2 \mrm{Im} \, \tilde{\Sigma}_g (x)
  \ne 0$ for $0 < |\kappa| < \kappa_{\ast}$. Therefore,  $\tilde{F}_{\kappa}$ has no zeros in $[\nu -r , \nu +r]$. Consequently, $\tilde{F}_{\kappa}$ has a unique zero in  $\overline{D}_{r}(\nu) \cap \mbb{C}_{-}$.    $\square$

\section{Examples}
In this section, we apply the main theorem to concrete examples of the 
Friedrichs--Lee Hamiltonian. Let $\msc{K} = L^2(\mbb{R}^d)$ with $d \in \mbb{N}$, 
and let $S=\omega$ be the multiplication operator by $\omega(k) = \sqrt{|k|^2 + m^2}$ 
with $m \geq 0$. When $m > 0$, we assume $\nu > m$, and when $m = 0$, 
we assume $\nu > 0$. We show that the main theorem applies by considering the cases $m > 0$  and $m = 0$ separately.

\subsection{Case $m > 0$}
Assume $m > 0$. We see that
$ \Hk - m \one  = 
\begin{pmatrix} \tilde{\nu} & \kappa A(g) \\ 
\kappa A^{\dagger}(g) & \tilde{S} \end{pmatrix} $, 
where $\tilde{\nu} = \nu - m$ and $\tilde{S} = S - m$. 
Since $\tilde{S} \psi (k)=\left( \sqrt{|k|^2 + m^2} - m\right) \psi (k)$ for $\psi \in \msc{D}(\omega )$, $\tilde{S}$ satisfies (A.1) and (A.3). 
Let $\msc{E}$ be the set of all entire functions on $\mbb{C}$. 
Let
$$
\msc{A} = \left\{ \phi(k) = |k|^{\frac{4-d}{2}} u(|k|^2) \;\middle|\; 
u \in \msc{E}, \int_{0}^{\infty} x |u(x)|^2 \, dx < \infty \right\}.
$$
For a spherically symmetric function $f$ on $\mbb{R}^d$, it holds that 
$ \int_{\mbb{R}^d} f(x) \, dx = S^{d-1} \int_0^\infty f(r) \, r^{d-1} \, dr
$ with $r = |x|$, where $S^{d-1}$ denotes the surface area of the unit sphere in $\mbb{R}^d$ with $S^{0}=2$.
For $\phi \in \msc{A}$, the polar coordinate transformation $|k|=r$ and the change of variables $x= r^2 $ yield
$$
\int_{\mbb{R}^d} |\phi(k)|^2 \, dk 
= \frac{S^{d-1}}{2} \int_{0}^{\infty} x |u(x)|^2 \, dx < \infty.
$$
Thus, $\msc{A} \subset L^2(\mbb{R}^d)$ holds. 
Next, we find the spectral measure $\mu^{\tilde{S}}_{\phi, \psi}$ 
and its density function $\rho^{\tilde{S}}_{\phi, \psi}$ for 
$\phi, \psi \in \msc{A}$. Let $\phi(k) = |k|^{\frac{4-d}{2}} u_1(|k|^2)$ 
and $\psi(k) = |k|^{\frac{4-d}{2}} u_2(|k|^2)$. 
For a Borel set $J \subset [0, \infty)$, we have
\begin{align}
\mu^{\tilde{S}}_{\phi, \psi}(J)
&= S^{d-1} \int_{\{ r \in [0, \infty) \mid \sqrt{r^2 + m^2} - m \in J \}} 
   r^3 \overline{u_1(r^2)} u_2(r^2) \, dr  \notag \\
&= S^{d-1} \int_{J} (\lambda^2 + 2m\lambda) (\lambda + m) 
   \overline{u_1(\lambda^2 + 2m\lambda)} u_2(\lambda^2 + 2m\lambda) \, d\lambda. \label{11/22a1}
\end{align}
From (\ref{11/22a1}), the density function $\rho^{\tilde{S}}_{\phi, \psi}$ is given by
\begin{equation}
\rho^{\tilde{S}}_{\phi, \psi}(\lambda) 
= S^{d-1} (\lambda^2 + 2m\lambda) (\lambda + m)  \overline{u_1(\lambda^2 + 2m\lambda)} u_2(\lambda^2 + 2m\lambda) 
   \quad \left( \lambda \in [0, \infty) \right). \label{11/22a2}
\end{equation}
We now  consider a specific function $g \in \msc{A}$ defined by 
$$ 
g(k) = |k|^{\frac{4-d}{2}} e^{-a |k|^2} \qquad (a>0) .
$$
Let us check (A.2).  Since $g \in \msc{A}$, $g \in L^2 (\mbb{R}^d)$.
It is seen that for all $n \in \mbb{N} $,
\begin{equation}
\int_{\mbb{R}^d} |k|^{2n} |g(k)|^2 \, dk = 
 S^{d-1} \int_{0}^{\infty} r^{2n + 3} e^{-2a r^2} \, dr < \infty .
\label{11/12a3}
\end{equation}
Then, setting $n = 1$ in (\ref{11/12a3}), we have $\| S g \|< \infty $. 
Since $\| \tilde{S} g \| \leq \| S g \| + m \| g \|$, we have $ g \in \msc{D} (\tilde{S})$. 
Moreover, we have 
$$ 
\int_{\mbb{R}^d} \frac{|g(k)|^2}{\sqrt{|k|^2 + m^2} - m} \, dk 
= S^{d-1} \int_0^\infty r \left( \sqrt{r^2 + m^2} + m \right) e^{-2a r^2} \, dr < \infty. 
$$
Thus, $g \in \msc{D}(\tilde{S}^{-1/2})$, and hence $\tilde{S}$ and $g$ 
satisfy (A.2). Next, we check (A.4) and (A.5). 
From (\ref{11/22a2}), $\rho^{\tilde{S}}_{g}$ is given by
$$
\rho^{\tilde{S}}_{g}(\lambda) = S^{d-1}  
(\lambda^2 + 2m\lambda) (\lambda + m) e^{-2a (\lambda^2 + 2m\lambda)} 
\quad \left( \lambda \in [0, \infty) \right).
$$
We see that $\rho^{\tilde{S}}_{g}$ is continuous on $[0, \infty)$ and 
satisfies $\rho^{\tilde{S}}_{g}(\lambda) > 0$ for all $\lambda > 0$. 
Hence, (A.4) is satisfied. Furthermore, for all $0 < \delta < \tilde{\nu}$, 
$\rho^{\tilde{S}}_{g}$ is analytic on  $I_{\delta} = (\tilde{\nu} - \delta, \tilde{\nu} + \delta)$ and has
an entire analytic continuation, and in particular an analytic continuation to $D_{\delta}(\tilde{\nu})$ given by
$$
\tilde{\rho}^{\tilde{S}}_{g}(z) = S^{d-1} (z^2 + 2mz) (z + m) 
e^{-2a (z^2 + 2mz)}.
$$
Therefore, (A.5) is also satisfied.  Let us check (A.6).  By (\ref{11/22a2}) we see that $\rho^{\tilde{S}}_{\phi, \psi}$, 
$\rho^{\tilde{S}}_{\phi, g}$ and $\rho^{\tilde{S}}_{g, \psi}$ have  entire analytic continuations, respectively. Thus, for all $\phi, \psi \in \msc{A}$,  (A.6) is satisfied.

\subsection{Case $m=0$}
Since $\omega(k) = |k|$, (A.1) and (A.3) are satisfied. 
Similarly to the case $m > 0$, we define a set $\msc{A}$ by
$$
\msc{A} = \left\{ \phi(k) = |k| u(|k|^2) \;\middle|\; 
u \in \msc{E}, \int_{0}^{\infty} x^{\frac{d}{2}} |u(x)|^2 \, dx < \infty \right\} .
$$
For all $\phi \in \msc{A}$,  we have
$$
\int_{\mbb{R}^d} |\phi(k)|^2 \, dk 
= \frac{S^{d-1}}{2} \int_{0}^{\infty} x^{\frac{d}{2}} |u(x)|^2 \, dx < \infty .
$$
Thus, $\msc{A} \subset L^2(\mbb{R}^d)$ holds. 
Next, we find the spectral measure $\mu^{S}_{\phi, \psi}$ and its density function $\rho^{S}_{\phi, \psi}$ for $\phi, \psi \in \msc{A}$ with  $\phi(k) = |k| u_1(|k|^2)$ and $\psi(k) = |k| u_2(|k|^2)$. 
For a  Borel set $J \subset [0, \infty)$, we have
\begin{equation}
\mu^{S}_{\phi, \psi}(J)
= \int_{\{ k \in \mbb{R}^d \mid |k| \in J \}} 
   |k|^2 \overline{u_1(|k|^2)} u_2(|k|^2) \, dk = S^{d-1} \int_{J} \lambda^{d+1}     \overline{u_1(\lambda^2)} u_2(\lambda^2) \, d\lambda . \label{11/22b1}
\end{equation}
Therefore, the density function $\rho^{S}_{\phi, \psi}$ is given by
\begin{equation}
\rho^{S}_{\phi, \psi}(\lambda) = S^{d-1} \lambda^{d+1} 
\overline{u_1(\lambda^2)} u_2(\lambda^2) 
\quad \left( \lambda \in [0, \infty) \right) . \label{11/22b2}
\end{equation}
Let $g \in \msc{A}$ be a specific function defined  by 
$$ 
g(k) = |k| e^{-a |k|^2} \qquad (a>0).
$$
Let us check (A.2). Since $g \in \msc{A}$, $g \in L^2 (\mbb{R}^d)$. For  $\beta \geq -\frac{1}{2}$, we have 
\begin{equation}
\int_{\mbb{R}^d} |k|^{2\beta} |g(k)|^2 \, dk = 
\frac{S^{d-1}}{2} \int_0^\infty x^{\beta + \frac{d}{2}} e^{-2ax} \, dx < \infty . \label{11/22b3}
\end{equation}
Setting $\beta = 1, -\frac{1}{2}$ in (\ref{11/22b3}), we have $g \in \msc{D}(S) \cap \msc{D}(S^{-1/2})$. Hence (A.2) is satisfied. 
Next, we check (A.4) and (A.5). From (\ref{11/22b2}), the density function $\rho^{S}_{g}$  is 
$$
\rho^{S}_{g}(\lambda) = S^{d-1} \lambda^{d+1} e^{-2a \lambda^2} 
\quad \left( \lambda \in [0, \infty) \right) .
$$
It is seen that $\rho^{S}_{g}$ is continuous on $[0, \infty)$ and 
satisfies $\rho^{S}_{g}(\lambda) > 0$ for all $\lambda \in (0, \infty)$. 
Hence, (A.4) is satisfied. Let $0 < \delta < \nu$. We see that 
$\rho^{S}_{g}$ is analytic on $I_{\delta} = (\nu - \delta, \nu + \delta)$ 
and has an entire analytic continuation, and in particular an analytic continuation to $D_{\delta}(\nu)$ given by
$$
\tilde{\rho}^{S}_{g}(z) = S^{d-1} z^{d+1} e^{-2a z^2} .
$$ 
Hence, (A.5) is satisfied.   Let us check (A.6). 
From  (\ref{11/22b2}), it follows that  
$\rho^{S}_{\phi, \psi}$, $\rho^{S}_{\phi, g}$ 
and $\rho^{S}_{g, \psi}$ have entire analytic continuations, respectively. Thus, for all $\phi, \psi \in \msc{A}$, (A.6) is satisfied.

$\;$ \\
{\large {\textbf{Acknowledgements}}} \\
The authors would like to thank Professor Fumio Hiroshima for his valuable comments and advice.



\begin{thebibliography}{99}
\small{
\bibitem{AgCo71}
J. Aguilar and J. M. Combes, A class of analytic perturbations for one-body Schr\"{o}dinger Hamiltonians, \textit{Commun. Math. Phys.} \textbf{22}, 269--279, (1971).
 \bibitem{Arai05}
A. Arai, \textit{Mathematical principles of quantum phenomena}, Asakura Shoten, 2005.  (in Japanese)
 \bibitem{BFS99} 
   V. Bach, J. Fr\"{o}hlich and I.M. Sigal,  Spectral analysis for systems of atoms and molecules coupled to the quantized radiation field, \textit{Commun. Math. Phys.}  \textbf{207}, 249--290, (1999).
\bibitem{BaCo71}
   E. Balslev and J. M. Combes, Spectral properties of many-body Schr\"{o}dinger operators with dilatation-analytic interactions, \textit{Commun. Math. Phys.} \textbf{22}, 280--294, (1971).
   \bibitem{FLL21}
    P. Facchi, M. Ligab\`{o} and D. Lonigro, 
   Spectral properties of the singular Friedrichs--Lee
Hamiltonian, \textit{J. Math. Phys.} \textbf{62}, 032102, (2021).
\bibitem{Fr48}
K.O. Friedrichs, On the perturbation of continuous spectra, \textit{Comm. Pure Appl. Math.} \textbf{1}, 361--406, (1948).
\bibitem{GP11}
M. Gadella and G.P. Pronko, The Friedrichs model and its use in resonance phenomena, \textit{Fortschr. Phys.}  \textbf{59}, 795--859, (2011).
\bibitem{HiSi96}
P.D. Hislop and I.M. Sigal, \textit{Introduction to spectral theory: With applications to Schr\"{o}dinger operators}, Springer, 1996.
\bibitem{King91}
C. King, Exponential decay near resonance without analyticity, 
\textit{Lett. Math. Phys.},  \textbf{23}, 215--222, (1991).
\bibitem{Lee54}
T.D. Lee, Some special examples in renormalizable field theory, \textit{Phys. Rev.} \textbf{95}, 1329--1334, (1954).
 \bibitem{ReSi78}
M. Reed and B. Simon, \textit{Methods of Modern Mathematical Physics Vol.IV: Analysis of operators}, Academic Press, 1978.
}
\end{thebibliography}
\end{document}